\documentstyle[12pt]{article}
\newlength{\pubnumber} 

\catcode`\@=11
\@addtoreset{equation}{section}
\def\section{\@startsection{section}{1}{\z@}{3.5ex plus 1ex minus .2ex}
 {2.3ex plus .2ex}{\large\bf}}
\def\subsection{\@startsection{subsection}{2}{\z@}{2.3ex plus .2ex}
 {2.3ex plus .2ex}{\bf}}

\begin{document}
\begin{titlepage}
\samepage{
\setcounter{page}{1}
\vfill
\begin{center}
 {\Large \bf Substructure of String\footnote
{Supported in part by
the Department of Energy under grant DE-FG05-86ER-40272,
and by the Institute for Fundamental Theory.}}
\vfill
 {\large Charles B. Thorn\footnote{E-mail  address: thorn@phys.ufl.edu}
   \\}
\vspace{.12in}
 {\it  Institute for Fundamental Theory\\
Department of Physics, University of Florida, Gainesville,
FL, 32611, USA}
\end{center}
\vfill
\begin{abstract}
I review work developing the idea that string is a composite
of point-like entities called string bits. Old and new insights
this picture brings into the nature of string theory are
discussed. This paper summarizes my
talk presented to the Strings96 conference at 
Santa Barbara, CA, 14-20 July 1996.
\noindent \end{abstract}
\vfill
\smallskip}
\end{titlepage}
\def\beq{\begin{equation}}
\def\eeq{\end{equation}}
\def\beqn{\begin{eqnarray}}
\def\eeqn{\end{eqnarray}}
\hyphenation{su-per-sym-met-ric non-su-per-sym-met-ric}
\hyphenation{space-time-super-sym-met-ric}
\hyphenation{mod-u-lar mod-u-lar--in-var-i-ant}
\setcounter{footnote}{0}
\def\balpha{\mbox{\boldmath$\alpha$}}
\def\bgamma{\hbox{\twelvembf\char\number 13}}
\def\bsigma{\hbox{\twelvembf\char\number 27}}
\def\bepsilon{\hbox{\twelvembf\char\number 15}}
\def\sgone{{\cal S}_1{\cal G}}
\def\sgtwo{{\cal S}_2{\cal G}}
\def\ob{\overline}
\def\bx{{\bf x}}
\def\by{{\bf y}}
\def\bz{{\bf z}}
\def\D{{\cal D}}
\def\R{{\cal R}}
\def\Q{{\cal Q}}
\def\G{{\cal G}}
\def\O{{\cal O}}
\def\Nlarge{N_c\rightarrow\infty}
\def\Tr{{\rm Tr}}
\def\ie{{\it i.e.}\ }
\def\etal{{\it et al.}}
\def\dash{---}
\newcommand{\goesas}[2]{\ \ {\phantom{a} \atop 
                \widetilde{#1\rightarrow #2}}}
\newcommand{\ket}[1]{|#1\rangle}
\newcommand{\bra}[1]{\langle#1|}
\newcommand{\firstket}[1]{|#1)}
\newcommand{\firstbra}[1]{(#1|}
\section{Introduction}
Today I would like to describe work 
which attempts to develop a theory of string based on the idea
that string is made of smaller 
entities\cite{thornmosc,bergmantbits,bergmantoned,klebanovs}. 
I have tried to 
encapsulate these ideas in a basic hypothesis, about the nature
of string, which I believe will survive even if our particular
efforts to implement it falter:
\begin{description}
\item [Hypothesis:] String and related structures in
{\bf $d+2$} space-time dimensions are {\bf composites} of
point-like entities, {\bf string bits}, which move in {\bf
at most $d$ space dimensions}.
\end{description}
I have emphasized in bold type the crucial aspects of the hypothesis.
If it is true, it should be possible to formulate string
theory in 4 dimensional space-time as a theory of a world
with only 2 space dimensions. If string theory is the right
description of nature, it would be a concrete realization of
`t Hooft's idea that the world is a hologram\cite{thooftbh}, an idea also
vigorously pursued by L. Susskind\cite{Susskindbh}.

It will be characteristic of an underlying theory, of the type
envisaged above, that it will not in any obvious way describe
gravity. How could it? The space or manifold on which the geometry
of gravity could be defined is not completely present in the formulation.
String bit dynamics must accomplish two rather remarkable
feats:
\begin{itemize}
\item It must produce {\bf at least} one new spatial
dimension. Moreover, any new manufactured dimensions
must behave as though they were on exactly the same footing
as the spatial dimensions already present in the underlying
theory.
\item Having produced new dimensions, the dynamics
must {\bf induce gravity} in this higher dimensional space.
\end{itemize}

It is this last point that offers the possibility of avoiding
information loss paradoxes. Since gravity will be an induced
collective effect, there is no obstruction to insisting that
the underlying theory be {\bf unitary}. In effect, we have turned
the problem around: rather than trying to establish that
{\bf the} theory of gravity is or is not consistent with unitarity,
we ask whether gravitational phenomena can be induced in
a unitary theory. I hasten to add that there is, of course, 
no a priori guarantee of a positive outcome.

Having agreed to build an underlying theory of string with at
least one missing dimension, we next must decide how much symmetry
the bit dynamics should retain. There is no single answer. However
for definiteness we shall provisionally retain a {\bf maximal}
symmetry consistent with our hypothesis. We shall assume there
is only one missing space dimension. The space-time boost
group should be a maximal subgroup, acting on the smaller space, 
of the Poincar\'e group. 
There are two possibilities.
I suppose most physicists would choose the $O(d,1)$ subgroup
of Poincar\'e $O(d+1,1)$. However, an equally logical choice
would be Galilei$(d,1)$. We choose the latter for reasons of
simplicity and expediency. I include expediency because Galilei invariant
dynamics is just that of ordinary non-relativistic quantum 
mechanics. Making this choice allows us to tap all of the many
methods and techniques developed by our condensed matter colleagues
over the years to deal with diverse physical phenomena. 
These resources will aid the implementation of our 
hypothesis, which must effectively handle
composite systems and collective excitations.

Finally, we come to supersymmetry. Once we choose Galilei as our
boost group, we are forced to incorporate Galilei supersymmetry.
It turns out that the maximal Galilei superalgebra is very
close in structure to the Poincar\'e superalgebra that would
be present in the higher dimensional theory. There are two types
of supercharge: $Q^A$, which is a square root of the
Newtonian mass, and $R^{\dot A}$, which is a square root of the
Hamiltonian $H$. The Newtonian mass of a system of string bits is just
$mM$, where $m$ is the mass of a bit, and $M$ is the number of bits.
The maximal Galilei superalgebra is
\begin{eqnarray}
 \{Q^A,Q^B\} = mM\delta^{AB}\; &,& \;
 \{Q^A,R^{\dot B}\} = {1\over 2}\balpha^{A\dot B}\cdot {\bf P}\; ,
    \nonumber\\
  \{R^{\dot A},R^{\dot B}\} &=& \delta^{{\dot A}{\dot B}}H/2 \; ,
\label{twoplusonealg}
\end{eqnarray}
%
where the components of $\balpha$ 
are elements of a $d$ dimensional Clifford algebra and ${\bf P}$
is the total $d$ dimensional momentum. Notice that this would
become precisely the $O(d+1,1)$ Poincar\'e superalgebra with
the replacement of $mM$ by $P^+$ and $H$ by $P^-$!
I must concede here that the last of these equations, the closure
of $R$ onto the Hamiltonian, is technically difficult to implement.
In the models Bergman and I have so far considered, it fails
for higher dimensional models: there are extra terms on the r.h.s.
not proportional to the Kronecker delta. For $d=1$, the supercharges
have only one component each and the full algebra is satisfied
by default. Studying this $d=1$ ``toy'' model 
has led to some important insights into
the behavior of superstring bit models. We still believe 
higher dimensional models with
the maximal supersymmetry can be constructed.

\section{Dynamics of String Bits}
The dynamics we set up for string bits is guided by the way string
dynamics works in light-cone gauge. Light-cone coordinates are
introduced in $d+2$ space-time dimensions by defining
$x^\pm=(t\pm x^{d+1})/\sqrt{2}$ and by denoting the remaining
$d$ ``transverse'' components ${\bf x}$. The corresponding
momentum components are $P^\pm$ and ${\bf P}$. Notice that $P^-$
is conjugate to $x^+$ and $P^+$ is conjugate to $x^-$. 
The relativistic dispersion law for a single particle state with
rest mass $\mu$  then reads
$$P^-={{\bf P}^2+\mu^2\over2P^+}.$$
Light-cone gauge for a free string is specified by
identifying the time-like world sheet parameter $\tau$ with
$x^+$ and choosing $\sigma$ so that ${\cal P}^+$, the density
of $P^+$ is uniform along the string. In this gauge the dynamics
is precisely that of a non-relativistic string in $d$ dimensions.

The coordinate ${\bf x}$ of a string bit has $d$ components--just like the
transverse coordinates of the light-cone. It possesses
a Newtonian mass $m$, and can carry momentum ${\bf p}$. 
The new dimension is generated by considering objects
made of a large variable number $M$ of bits. Then $mM$ can be
interpreted as an effectively continuous $P^+$. 
There is a dynamical requirement
for this interpretation to work. By Galilean invariance one can always
separate the center of mass motion so that
we automatically have for any collection of particles
$$H={{\bf P}^2\over2mM}+H_{int},$$
where ${\bf P}$ is the total momentum and $H_{int}$ is the energy
of internal motion. To achieve the relativistic dispersion law
with $P^+=mM$, it must be the case that the excitation energies
associated with the internal motion scale as $1/M$ for large $M$.
This behavior can be expected for a wide class of one dimensional
composite systems, but it is by no means guaranteed.
\section{Insights from the Composite Picture of String}
\subsection{String as a Polymer of String Bits} 
  To get even a noninteracting string we must confront a many body
bound state problem. The bits must interact with each other in
a way that allows them  bind into arbitrarily long polymers.
It is not hard to set up such a dynamics if we imagine that the
bits have somehow been ordered around a loop. Then a Hamiltonian
$$H={1\over2m}\sum_{k=1}^M{\bf p}_k^2-{1\over m}\sum_{k=1}^M
{\cal V}({\bf x}_{k+1}-{\bf x}_k)$$
will predict polymer formation provided the potential
$-{\cal V}$ is attractive enough to bind a pair of bits. If so the
bond breaking energy $B=O(1/m)$. The low energy internal excitations will
correspond to vibrational frequencies $\omega_n/m$. One can show
that these frequencies scale as $n/M$, so that a string tension
$T_0$ can be defined by
$$\omega_n\goesas {\scriptstyle M}\infty {2\pi T_0n\over M}.$$
This scaling law ensures the relativistic dispersion law.
Notice that the energy to break a bond is huge compared to the
excitation energies for large $M$: the bits are effectively confined to 
polymers.
\subsection{Issues of Stability}
The dynamics of the polymer system we considered 
in the previous section was
artificial in that the cyclic ordering of bits around a loop was
preserved. Any realistic many body dynamics would permit the bonding
pattern to be rearranged. Such rearrangements would include ones
in which the number of polymers changes. A polymer
in its ground state could be unstable to decay into two smaller 
polymers. To address this issue compare the  energy
of an $M$ bit polymer at rest in its ground state to the
energy of a system of two polymers, one with $M_1<M$ and the
other with $M-M_1$ bits, each at rest in their ground states.
The ground state of each polymer is determined by the
Hamiltonian of the previous section. By the uncertainty
principle the internal kinetic energy of the two polymer
system is less than that of the one polymer system, since the
latter is more localized. Because there are exactly $M$ bonds 
in both systems, and no non-nearest neighbor interactions, the
two polymer system will be more tightly bound (since it
has less internal kinetic energy), so its potential energy
will also be less. Thus we conclude that
$$E_G(M)>E_G(M_1)+E_G(M-M_1).$$
Since the large $M$ behavior of $E_G$ is expected to have the
asymptotic form
$$E_G(M)\goesas{\scriptstyle M}\infty\alpha M + {\gamma\over M} + O({1\over
M^3}),$$
we conclude that $\gamma<0$. But $2\gamma$ is just the (mass)${}^2$
of the lowest string state, \ie\ it is a tachyon. This provides
a clear understanding of the tachyonic instability of the bosonic
string\cite{thornweeparton}. 
It seems unavoidable if the only bit dynamical variables are
${\bf x}$ and ${\bf p}$.
\subsection{Bits Can Have Discrete Degrees of Freedom.}
Of course bits can also occur in multiplets. For example,
a bit could carry spin or ``flavor'' such as isospin $1/2$. When 
such bits are
bound into polymers, spin or isospin waves will naturally
describe some of the low energy excitations. Isospin waves have precisely
the energy and degeneracy patterns of a compactified 
coordinate\cite{gilesmt}. Spin waves would provide the world sheet fields
$H^\mu$, $\Gamma^\mu$ of the NS, respectively R sectors of the spinning
string. The contributions of spin or isospin wave fluctuations to 
the ground state energy of the polymer depend on the bit
number. If $M$ is even (NS sector), the effective $\gamma$ is
negative, but $\gamma$ is positive for odd $M$. That doesn't
really help with stability, since a polymer with $M$ odd
can decay into two polymers, one of which must have even
bit number. We know of course, that in the NSR formulation
of superstring theory, the closed string tachyon can be
consistently projected out of scattering amplitudes through
the miracle of the GSO projection. We would prefer that
a bit model for superstring not depend on such a delicate
cancelation. However, the possibility of interpreting
compactified dimensions as ``flavor waves'' remains
an intriguing way to manufacture the extra dimensions
required by critical superstring.
\subsection{Statistics Waves}
If string bits occur in multiplets which contain both
fermions and bosons, fluctuations in statistics would naturally
lead to what we call {\bf statistics waves}\cite{bergmantoned}. 
Unlike spin or isospin waves, the $\gamma$ 
parameter for statistics waves is positive
regardless of whether $M$ is even or odd. Thus statistics
fluctuations can stabilize large polymers. When bits are put
in supersymmetry multiplets, these waves on large polymers
are precisely described by the Green-Schwarz spinor valued
world sheet fields. For superstring, the contribution to the
energy of
statistics fluctuations exactly cancels that of vibrational
fluctuations, \ie $E_G=0$ {\bf exactly}. Thus all super polymers, 
large and small are
marginally stable. 
\subsection{Interacting Polymers}
 To handle polymers that can interact with each other we must specify
a complete many body dynamics, which doesn't artificially suppress
bond rearrangement. 
We give each bit an adjoint ``color'' degree of freedom by
introducing a super-bit creation operator that is a matrix transforming
in the adjoint representation of color $SU(N_c)$. Quite generally,
$$ \phi^{\dagger f\beta}_{a_1\cdots a_n \alpha}({\bf x})$$
will create a bit at the position ${\bf x}$, with flavor $f$, spin
state specified by $a_1\cdots a_n$ (bosons (fermions if $n$
is even (odd)), and color $\alpha,\beta$ where
lower (upper) color indices transform in the $N_c$ (${\bar N}_c$)
representations of $SU(N_c)$. We can employ `t Hooft's $1/N_c$ 
expansion\cite{thooftlargen} to 
isolate a limit ($N_c\rightarrow\infty$) in which the bits
on a polymer maintain their cyclic order and polymers do not interact
with each other\cite{thornfock}. The Hamiltonian is very complicated,
but if we suppress indices, it has roughly the schematic form:
$$H=\int d{\bf x}{1\over2m}\Tr[\nabla\phi^\dagger\cdot\nabla\phi] 
+{1\over mN_c}\int d{\bf x}d{\bf y}{\cal V}({\bf x}-{\bf y})
\Tr[\phi^\dagger({\bf x}),\phi({\bf x})][\phi^\dagger({\bf y}),\phi({\bf y})].
$$
At leading order in $1/N_c$ a state of the form
\begin{eqnarray}
 \ket{\Psi(&x_1&,\ldots,x_M)} =\int dx_1\cdots dx_M \nonumber\\
 &&~~~~~~\times\Tr[\phi^\dagger(x_1)\cdots\phi^\dagger(x_M)]\ket{0}
     \Psi(x_1,\ldots ,x_M)\; ,
 \label{chain}
\end{eqnarray}
can be shown to be an energy eigenstate with the properties
of a noninteracting  closed polymer.
At order $1/N_c$ this Hamiltonian allows bond rearrangement in
which a single closed polymer can transform into two closed polymers.
The amplitude for such a process can be expected to have the same
large $M$ scaling behavior as the discretized closed string vertex
studied in \cite{thornlcft}. Assume the two polymers in the final
state have bit numbers $fM$ and $(1-f)M$. Then for the bosonic
string this vertex was found to 
have the large $M$ behavior
$${\rm Vertex} \sim {1\over N_c}
\left({M\over m}\right)^{3/2}\left({1\over M}\right)^{d/8}\rightarrow
{1\over N_c}\left({1\over P^+}\right)^{(d-12)/8}m^{(d-24)/8}, $$
where the $d$ independent factor is due to wave function normalization and
the $d$ dependence comes from the overlap of initial and final wave functions.
There are three distinct cases. For $d<24$ the amplitude blows up
in the continuum limit $m\to0$. This is the sub-critical string.
By taking a double scaling limit, $N_c\to\infty$, as $m\to 0$
a finite limit can be obtained, but Lorentz invariance is lost.
The case $d=24$ is the critical string and the vertex is finite
and consistent with Lorentz invariance (${1/ P^+}^{3/2}$). 
Finally, for $d>24$ the
vertex vanishes and the theory is trivial.
\subsection{Why Bit Dynamics Must Sometimes Imply Gravity}
If bit dynamics is to induce gravity, it must induce
a Poincar\'e invariant higher dimensional theory which includes a massless
spin 2 particle (the graviton) in its spectrum {\bf as interpreted
in the effectively higher dimension theory}. This is a tall order.
However, as argued by Mueller\cite{mueller}, polymers of
string bits generically scatter
in a certain kinematic region in a way precisely consistent with
this happening. Actually his argument was applied to light cone
string, but it applies equally to our polymers. Consider forward
scattering of a polymer of bit number $M_1$ off a polymer of
bit number $M_2$. Both $M_1$ and $M_2$ are assumed large (of $O(1/m)$),
but $M_2>>M_1$. Then the energy of 2 is much less than that of 1
(remember $H\sim 1/mM$), which is of $O(1)$. The total energy
of the system is of $O(1)$, so the interaction time of the
polymers will be of order $1/E=O(1)$ The small polymer will
strike the large polymer at any of $M_2$ locations.
But because of the finite
velocity of sound on the large polymer, a small polymer must be
emitted from a location not very distant from the initial
impact, because the emission must occur within a time (either
before or after the impact) of
order O(1). As $M_2\to\infty$ only a relatively tiny part of the
large polymer is disturbed, so that the probability of 
scattering for each impact must approach a constant, and
since there are $M_2$ possible impact locations, we conclude that
$${\rm Cross~section}\goesas{\scriptstyle M_2}\infty{\rm const}\times M_2.$$

What does this have to do with gravity? Not much, unless one
{\bf knows} that the physics predicted by the bit
model turns out to be Poincar\'e invariant. Not all string bit models have
this property--a simple counterexample being the sub-critical
bosonic model. But if Poincar\'e invariance does hold with
the identification $P^+=mM$, then, for the kinematics
considered above the Mandelstam invariant
$$s=2(P_1+P_2)^+(P_1+P_2)^--({\bf P}_1+{\bf P}_2)^2\approx
2mM_2E_1$$
is large and the momentum transfer invariant $t=0$ because
forward scattering is being considered. This is just the
kinematics of Regge behavior in which the cross section
should have the behavior
$$\sigma\sim Ks^{\alpha(0)-1},$$
where $\alpha(t)$ is the highest lying Regge trajectory with
vacuum quantum numbers. Comparison to the linear growth following from
Mueller's argument then implies that 
$$\alpha(0)=2, \qquad{\rm if~scattering~is~Poincar\acute{e}~invariant},$$
Since the vacuum Regge trajectory has even signature, 2 is
a right signature point, so a massless spin 2 particle must
lie on this trajectory. This particle must also couple
at low momentum transfer, so that Weinberg's
soft graviton theorems will require couplings consistent with
general covariance. Thus Mueller's argument together with
Poincar\'e invariance would require that gravitational
phenomena be induced.

 To understand what happens when Poincar\'e invariance is {\bf not}
obtained, consider the sub-critical bosonic model. In that case
the 4 polymer cross section has the scaling behavior
$$\sigma\sim\left({1\over M_2}\right)^{(d-12)/12}{\cal M},$$
where ${\cal M}$ is an integral over moduli space of an
integrand that continues smoothly from the direct to
crossed channel singularities. Thus it is the large $M_2$
behavior of ${\cal M}$ that is controlled by crossed
channel Regge trajectories. Then Mueller's argument predicts
$$\alpha(0)={d\over12}, \qquad{\rm for~subcritical~bosonic~model}.$$
This trajectory intercept implies that the spin 2 particle
lying on the trajectory is massive in accord with expectations
for the sub-critical case.
Of course, for $d=24$ the prediction reduces to the Poincar\'e
invariant result. 
\section{Size of Polymer Ground State}
In the previous section we enumerated several important 
insights string bit models provide into string dynamics.
All of those results have appeared in the literature, some
of them are many years old. In this section, I would like to 
briefly describe studies by Bergman and me on how
string interactions might affect the size of the polymer
ground state. The importance of this issue has been stressed
by L. Susskind.

It has been known for a long time that vibrational fluctuations
cause the size of the polymer ground state to grow
logarithmically with the bit number:
$$R^2_0\sim{1\over T_0}\ln M.$$
Although one might be disturbed that there is any growth
at all, Susskind has pointed out that the above growth is in
fact too slow to expect a perturbative description of the
ground state to have any validity at all. This is because the bit
number density will grow linearly (up to powers of
$\ln M$). Thus even though the string coupling might be
quite small, at large $M$ the density is so high that the interaction
energy is a factor $g^2M$ times the unperturbed energy.
Thus for $M>1/g^2$, interactions will dominate.

In our bit models the $g\sim 1/N_c$, so to examine the effects
of interactions, we must look at finite $N_c$ effects. There 
are many complications that arise. These include interaction
between non nearest neighbor bits on a polymer chain as well
as the possibility of rearrangements in the bond pattern.
These two effects arise together and it is not really
possible to separate them. Nonetheless, we thought it would
be interesting to first consider the effects of non-nearest
neighbor interactions, with the cyclic ordering of the bits
assumed fixed. Because the bit Hamiltonian is positive
definite, it can easily be seen that residual non-nearest
neighbor interactions should be repulsive. Thus we expected
them to increase the size of the ground state with an
accompanying decrease in bit density.

We studied a model in which the nearest neighbor attractive
interactions were supplemented by a short range repulsion
between all pairs of bits. For details, I refer the reader to
the poster session presented by Bergman \cite{bergmanposter}.
I present here only a summary of our results. We introduced
measures of size $R^2_k$, the mean squared distance between
bits separated by k stems along the chain. We also introduced an
overall size measure $R^2=(1/M)\sum_k R_k^2$. We used did
a variational calculation with a trial wave function given by
the ground state of a harmonic system characterized by
an arbitrary set of normal mode frequencies $\omega_n$ which
were our variational parameters. The main results are:
\begin{itemize}
\item $R_1^2$ stays finite at large $M$. This means that the
extra repulsions are not stressing the bonds excessively.
\item $R^2$ grows {\bf quadratically} with $M$. On the one
hand this means that the bit density is getting smaller,
not blowing up. On the other hand, it is discouraging, since
it is easy to show from dipole sum rules that 
$R^2<{1/ mE_{GAP}}.$
Thus quadratic growth of $R^2$ implies that $E_{GAP}$ the
energy gap between ground and first excited states is closing
faster than $1/M^2$, whereas a Poincar\'e invariant result
would entail a gap of order $1/M$. In fact,
\item $E_{GAP}\sim {1/ gM^3}$.
\end{itemize}
Although these results are discouraging, it must be remembered
that we are not really analyzing the string bit model itself
but rather a rough imitation of true bit dynamics. In particular
bond rearrangements have been disallowed, and that could be
quite important. We have also not attempted to study a 
supersymmetric model, and that is probably even more important.
Thus it is premature to draw a negative conclusion about
string bit ideas.
\section{Conclusion}
I hope to have convinced you that string bit models give 
important physical insight into the nature of string:
\begin{itemize}
\item Stability: the importance of statistics waves on polymers
of superbits.
\item The induction of gravity in a unitary theory.
\item Dimensional Enhancement: The promotion of Galilei$(d,1)$ to 
Poincar\'e$(d+1,1)$.
\end{itemize}
In addition these models provide a framework for addressing a
variety of questions unanswerable in string perturbation
theory.

As should be clear, there are many open problems and issues
that must be addressed. Probably most important is the
construction of fully interacting superstring bit models in higher than 1
space dimension in which the maximal Galilei superalgebra
closes. The question of whether interactions cause the size
of the polymer ground state to grow linearly with bit
number and not ruin the ``nice'' properties of the string
ground state is still open. An  improved numerical study
of these issues seems very feasible.


\end{document}